\documentstyle{article}
\newcommand{\f}{\begin{equation}}
\newcommand{\ff}{\end{equation}}

\begin{document}

 \vfil

 \centerline {{\bf The $G_{Newton} \rightarrow 0$
Limit of Euclidean Quantum Gravity}}
\medskip
\centerline{Lee Smolin}
\centerline{{\it Physics Department, Syracuse University}}
\centerline {{\it Syracuse, New York 13244-1130}}

\vskip 1in

\centerline{\bf Abstract}
\vskip.1in
\noindent

Using the Ashtekar formulation, it is shown that the $G_{Newton} \rightarrow
0$ limit of Euclidean or complexified general relativity  is not a
free field theory, but is a theory that
describes  a linearized self-dual connection propagating on an arbitrary
anti-self-dual background.  This theory is quantized in the loop
representation and, as in the full theory, an infinite dimnensional space of
exact solutions to the constraints are found.   An inner product is also
proposed.  The path integral is
constructed from the Hamiltonian theory and the measure is explicitly
computed  nonperturbatively, without relying on a semiclassical expansion.
This theory could provide the starting point for a new approach to
a perturbation theory in  $G_{Newton}$ that does not rely on a background
field expansion and in which full diffeomorphism invariance is satisfied at
each order.

\vfil
\rightline{\today}
\eject

\section{Introduction}

The starting point for all contemporary work on the problem of
quantum gravity
is the non-renormalizability of the standard perturbation theory which is
understood as an expansion, in powers of $l_{plank} = \sqrt{\hbar G/c^3}$,
of the
metric $g_{\mu \nu}$ around the flat metric $\eta _{\mu \nu}$. In
this semiclassical approach to perturbation theory, the quantum field
theory of gravitons is defined by making a shift in the definition
of the field operator
\f
\hat{g}_{\mu \nu} = \eta _{\mu \nu} + l_{Planck}\hat{h}_{\mu \nu}
\ff
and then constructing a quantum field theory for the {\it deviations}
$\hat{h}_{\mu \nu}$ from the fixed classical background $\eta _{\mu \nu}$.

Formulated this way, the $G_{Newton} \rightarrow 0$ limit of quantum gravity
is just a free field theory. However, in this standard formulation,
the limit $ G_{Newton} \rightarrow 0$ has two troublesome features.
First, the interpretation
of the gauge symmetry as diffeomorphism invariance is lost.
Second, properties
of the background spacetime which do not extend to the full theory, such as
its Poincare invariance, determine the operator ordering perscription used
in defining the free field theory around which the perturbation theory
is constructed.

The purpose of this paper is to show that if we do not
base the quantization on the shifted field (1) there is a different
$G_{Newton} \rightarrow 0$ limit of quantum gravity that, in the
Euclidean case, is not a free field theory.  It is instead an
interacting,  chirally asymmetric, theory that retains the exact
diffeomorphism invariance of the full theory.  The significance of
this result is that this formulation could be the starting point of a new
approach to perturbation theory that is completely quantum mechanical
in that it does not rely on a semiclassical expansion around a classical
background metric and so retains, at each
order, the full diffeomorphism invariance of general relativity.

More particularly, I will show first that, for the
the Euclidean or complexified classical theory, there is a
$G_{Newton} \rightarrow 0$ limit of general relativity that consists of
the full anti-selfdual, or left handed, sector of the theory,
together with a linearized $U(1)^3$ connection.  These latter fields
may be interpreted
as the linearization of the self-dual part of the Weyl tensor,
that propagates on
the anti-selfdual solution.   This chirally asymmetric result
follows from taking
the $G_{Newton} \rightarrow 0$ limit within the new-variables formulation of
Ashtekar\cite{abhay,book,poona}\footnote{For the {\it linearization} of
the Ashtekar formalism, which is not equivalent to the theory described
here, see \cite{gravitons}.}.
It shows that what $G_{Newton}$ measures is the coupling between the left
and right handed degrees of freedom of the theory.

This theory, which I will call the "googly" theory, as it is reminiscent
of the googly\cite{roger-googly}  problem of Penrose's
twistor program\cite{twistor},
is not a free field theory,
but it may be an integrable system.  As I will show in section 3 it
can be quantized
using the loop representation \cite{carlolee,gambini} which allows us to
construct exact solutions to the quantum constraints, as in the case of the
full theory.  Furthermore, in the case of the googly
theory, the quantization
overcomes several obstacles that are still unresolved in the
full theory, in that a larger set of
solutions to the constraints may be immediately obtained and a
finite dimensional
subalgebra of the physical observables can be found by inspection.

FInally, the paper closes with a new definition of the Euclidean path
integral in which the measure is determined nonperturbatively from the
canonical theory.  It is proposed that this integral, rather than
the free field
theory, should be the starting point for a perturbative analysis of
quantum gravity.

\section{The Classical Theory}

In the Ashtekar formulation, Newton's constant (which we will from now on
call $G$)
appears only in association with
the self-dual connection  $A_a$ as in
\f
F^i_{ab} = \partial_a A^i_b  -\partial_b A^i_a + G \epsilon^{ijk}A_a^j A_b^k,
\ \ \ {\cal D}_a \tilde{E}^b_i =
\partial_a \tilde{E}^{bk}
+ G \epsilon^{ijk}A_a^j\tilde{E}^{bk}
\ff
The googly theory is arrived at by simply setting $G =0$ in all such
expressions, while keeping $A_{ai}$ and $\tilde{E}^{bj}$
fixed.  (It will be useful to recall that
$A ^i_a$, has dimension of $(length)^{-3}$ while
$\hat E ^{ai}$, the conjugate momenta to $A^i_a$, is dimensionless.)
No assumption that $\hat E  ^{ai}$ is close
to a flat, or any other, background metric is made.

We may in fact begin by setting $G=0$ in the action for general relativity
in the self-dual form\cite{action}, which then becomes,
\f
S = \int \epsilon^{\mu \nu \alpha \beta} e_\mu^i e_\nu^j {}^4f_{\alpha
\beta ij}
\ff
where
${}^4f_{\mu \nu}^{ij} = \partial_\mu {}^4A_v^{ij} -
\partial_\nu {}^4A_\mu^{ij}$
and
the $e_\mu^i$ and ${}^4A_\mu^{i j} = - {}^4A_\mu ^{ji}$
are  $1$-forms on spacetime.
$A_\mu^{ij}$ is the self-dual connection, which satisfies
$A_\mu^{*ij} = {1\over 2}
\epsilon^{ijkl} A_\mu^{kl} = + A_\mu^{ij}$.
For the moment, all fields will be assumed
to be complex, reality conditions will be discussed shortly.

The Hamiltonian analysis may be performed as
usual\cite{action,book,poona,review}
on a three surface $\Sigma$,
leading to canonical variables $A_a^i$, which are three one forms on $\Sigma$,
and $\tilde E ^{aj}$, which are three vector densities on $\Sigma$, with
Poisson brackets,
\f
\{A_a^i (x), \tilde{E}^{bj} (y)\} = \delta_a^b \delta ^{ij}
\delta ^3 (x,y).
\ff
The constraints of the theory are
\f
g^i = \partial_a \tilde{E}^{ai}
\ff
\f
C_a = \tilde{E}^{ai} f^i_{ab}
\ff
\f
C= \epsilon^{ijk} \tilde{E}^{ai} \tilde{E}^{bj} f_{ab}^k
\ff
with $f_{ab}^k = 2\partial_{[a}A_{b]}^k$.  We may note that the $g^i$
generate a $U(1)^3$ internal gauge group, thus setting $G$ to zero may be
thought of as a contraction in which $SU(2) \rightarrow U(1)^3$.
The remaining constraint algebra is identical
to that of the Ashtekar formalism.  The equation of motion are then deduced
from the Hamiltonian
\f
H  = \int_\Sigma  N C  .
\ff
(we do not consider here boundary conditions or the boundary
terms which appear in the asymptotically flat case\cite{abhay,book,poona}.)
If we set $V^{ai} = N \tilde{E}^{ai} $ and choose  $N$ such that
$\partial_a N = 0$ the equations of motion  are
\f
\dot{V}^{ai} =  \epsilon^{ijk} [V^j , V^k ]^a
\ff

\f
\dot{A}_{ai} = \epsilon^{ijk}V^{jb} f^k_{ab}
\ff

Note that the $V ^{ai}$ decouple.  They are three divergence free vector
fields on $\Sigma$ which evolve according to (9).  This is exactly the
anti-self dual sector of the theory, as is easily seen from the fact that
$F_{ab}^i$, which is the self-dual part of the Weyl tensor,
vanishes\cite{selfdual}.  The $A_{a}^i$
then satisfy linear constraint
and evolution equations in the presence
of the anti-selfdual background $V^{ai}$ .
Because $f_{ab}^i= 2\partial_{[a} A _{b]}^i$ is the linearization
of the self-dual Weyl tensor we see that the $A_{ai}$ can be
interpreted as  the first order selfdual  fields propagating
on exact anti-selfdual metrics.  However, it should be emphasized that they
do not satisfy the linearization of Einstein's equations around the anti-self
dual backgrounds, setting $G=0$ in the Einstein equations is not the same
thing as linearizing them\footnote{I would like to thank
Abhay Ashtekar, Carlos Kozemeh, Carlo
Rovelli and Joseph Samuel for a discussion that clarified this point and
in particular for the remark that this result is not in conflict with the
theorem of Samuel\cite{remarkable}
on the non-existence of purely self-dual perturbations.}.

Note that all these equations  are well defined even when
$\tilde{E}^{ai}$ are
degenerate. Degenerate solutions to the full theory have been studied in
\cite{degenerate} and it is easy to see that
there are degenerate solutions in this case as well.
Around non-degenerate solutions the theory has $4$
phase space degrees of freedom, which can be seen by counting as well as by
expanding around flat space $\tilde{E}^{ai} = \delta^{ai} +  h^{ai}$.  We may
then fix symmetric traceless transverse gauge for $h^{ai}$ and $A_{ai}$ and
taking the fourier transform, following treatements of the
linearized theory\cite{book,gravitons} write the physical
degrees of freedom as,
\f
h^{ai} (p) = h^+ (p) m^a (p) m^i (p) + h^- (p) \bar{m}^a (p) \bar{m}^i (p)
\ff

\f
A_{ai}(p) = A^+(p) m^a (p) m^i (p) + A^- (p) \bar{m}^a(p) \bar{m}^i (p)
\ff
where the $m(p)^a$ are the complex polarization vectors that
satisfy $p^a m_a =0$, $m_a m^a = 0$ $m_a \bar{m}^a=1$.  From (9) and (10),
the  physical
modes evolve as,
\f
\dot{h}_\pm = \pm |p| h_\pm
\ff
\f
\dot{A}_\pm = \mp |p| A_\pm
\ff

So far, we have been using complex fields, to define the Euclidean or the
Minkowskian case we have to impose reality conditions.
To define the Euclidean theory, we impose the  conditions
\f
\tilde E^{ai} = \bar{\tilde{E}}^{ai}
\ff
\f
A _{ai} = \bar{A}_{ai}
\ff
so that in the nondegenerate case we have $4$ real
phase space degrees of freedom per point.  Full Euclidean general relativity
may now be constructed by a power series in $G$, by writing
\f
\tilde{E}^{ai} = \tilde{E}^{ai}_0 + G \tilde{E}^{ai}_1  + G^2
\tilde{E}^{ai}_2  + ...
\ff
\f
A_{ai} =A_{ai}^0 + G A_{ai}^1 + ...
\ff
and continuing the expansion of the constraints and equations of motion.

The situation is rather different in the Minkowskian theory
because there the reality
conditions involve $G$.  The Minkowskian realtity conditions
of the full theory
are that $\tilde{\tilde{q}}^{ab}$ and its time derivate should be
real.  The latter
condition can be expressed as,
\f
G (A_{ai} + \bar{A}_{ai}) = 2\Gamma (\tilde{E})_{ai}
\ff
where $\Gamma (\tilde{E})_{ai}$ is the three dimensional Christofel connection
of the frame fields $\tilde{E}^{ai}$\cite{abhay,book,poona}.
If we take $G$ to zero with $A_{ai}$ and
$\tilde{E}^{ai}$ held fixed we see that $\Gamma (\tilde{E})_{ai}=0$,
which means
the spatial geometry is flat.  This can be seen also by another route,
in which
we use the fact that the second reality condition (19) expresses the fact that
the time derivative of $\tilde{\tilde{q}}^{ab}$ should be real.  To work this
out we must recall that in the Minkowskian theory  the Poisson brackets (4)
and (9)
are multiplied by $\imath$ (or alternately, we take
$\tilde{\tilde{q}}^{ab}$ to be
negative definite so that the $\tilde{E}^{ai}$ are pure imaginary.)
We then find that
\f
{ d \tilde{\tilde{q}}^{ab} \over dt} = \imath N^{-2}
\epsilon_{ijk} [ V_j , V_k ]^{(a}V_i^{b)}
\ff
so that the requirement that this be real leaves us with
$\epsilon_{ijk} [ V_j , V_k ]^{(a}V_i^{b)}=0$, which implies, in
the nondegenerate
case, both that the metric
of three space is flat and that its time derivative is zero.
This is just the
well known fact that Minkowski spacetime is the only real
self-dual solution of
the Einstein equations.

The $A_{ai}$ are then not restricted by the Minkowskian reality conditions of
the $G \rightarrow 0$ limit of the theory we have considered here, and they
then carry four real phase space degrees of freedom.  We may note that these
dynamics are different from the {\it linearization} of the theory around
flat spacetime, in which  (19) couples the linear
fluctuations in $A_{ai}$ to the linear fluctuations in the metric around
a background\cite{gravitons}.  One way to see this is to consider a
different limit of the theory in which we first
expand  $\tilde{E}^{ai} = \delta^{ai} + G h^{ai}$, and then take the
$G \rightarrow 0$ limit.  We then see that to first
order the Minkowskian reality conditions mix $A_{ai}$ and $h_{ai}$.  For
the physical modes they are
\f
\bar{h} _{\pm}(p) = h _{\pm} (-p)
\ff

\f
G ( A_{\pm} (p) + \bar{A}_\pm (p) )= \mp 2 G |p| h_{\pm} (p)
\ff
Consistent expansion of the constraints, equations of motion and the reality
conditions in power of $G$ then reproduces the usual
perturbation expansion.

Finally, we may note that for
this googly theory it is immediately possible to write down physical
observables, which are also constants of motion for the compact case.  The
reader may verify that
\f
Q = \int_\Sigma \tilde E^{ai} A_{ai}
\ff
and
\f
Q^i = \int _\Sigma \epsilon^{ijk} \tilde E ^{aj}A_{ak}
\ff
commute with the constraints.  They generate the global scale transformations
and $SO(3)$ rotations, respectively.  It is interesting to note that
these are also observables for the opposite, {\it strong-coupling},
limit of the theory\cite{viqar-strong}.  If the theory is
integrable, it may be
possible to write down an explicit infinite dimensional algebra
of observables,
but this has not yet been done.

\section{ The Quantum Theory}

A natural set of variables to use to quantize the theory
are the loop observables, which are invariant under the internal
gauge transformations.  (For background and applications of the loop
representation,
see\cite{carlolee,gambini,review,gravitons,maxwell,loopym,2+1}.)
The $U(1)^3$  loop observables have already been studied
for linearized gravity\cite{gravitons}, they are:
\f
t^k[\gamma] \equiv e^{il^2\int A^k_a[\gamma (s)]\dot{\gamma}^a (s)ds}
\ff
where $l$ is a length which is $\underline {not}$ related to
 $G_{Newton}$, and is
only included because $A_a^i$ has dimensions of
$(length)^{-3}$.  We may note that, unlike the case of
linearized gravity\cite{gravitons},
in this model the frame fields {\it are} invariant under internal
gauge transformations.

The Poisson algebra is
\f
\{ \tilde E ^{ai} (x), t^k [\gamma ] \} =-l^2
\Delta^a [x, \gamma ] \delta^{ik} t^k [\gamma ]
\ff
where  $\Delta^a [x, \gamma ]  = \int ds \delta^3 (x,\gamma (s))
\dot{\gamma}^a (s)$.
The $t^k [\gamma ]$ satisfy the abelian loop relations,
\f
t^k [\alpha] t^k [\beta]  = t^k[\alpha \circ \beta ],
\ff
for loops $\alpha$ and $\beta$ with a common base point, where
$\circ$ is here the multiplication of loops.  Further, for the
Euclidean case we have,
\f
\overline{t^k [\gamma ] } = t^k [\gamma^{-1} ].
\ff
Finally,
we may note that $t^k [\gamma ]$ of a trivial loop is equal to $1$.

The quantization is performed on the triple nonparametric
loop space $({\cal HL}_1)^3$,
which is the same
space on which the loop representation for linearized gravity is
constructed\cite{gravitons}.  We first construct the nonparametric
loop space\cite{gambini} (which is also called the  $U(1)$
holonomic loop space) which is the space of sets of parametrized,
loops on $\Sigma$ on which
we have imposed the following equivalence relations:  i) invariance
under monotonic reparametrizations,  ii) for two loops $\alpha$ and
$\beta$ that have a common base point the set
$\alpha \cup \beta \approx \alpha \circ \beta $, iii) for a loop
$\alpha $ and an open line $\eta $ with a common base point
$\alpha \circ \eta \circ \eta^{-1} \approx \alpha $.  We call
this space ${\cal HL}_1$, as it incorporates all of the relations
among $U(1)$ holnomies.  We then consider the elements of
the product $({\cal HL}_1)^3$, which we will denote
$ \vec{\alpha} = \{\alpha_1, \alpha _2, \alpha _3\}$.
We then quantized the theory by defining  the (unphysical)
Hilbert space, $\cal S$, consisting of complex functions
$\psi [\vec{\alpha} ] $ on ${\cal HL}_1^3$ with the inner product
\f
<\Psi| \chi >\equiv \int d\mu [\alpha_1] d\mu [\alpha _2] d\mu [\alpha _3]
\bar{\Psi} [\vec{\alpha}] \chi [\vec{\alpha}]
\ff
where $d\mu [\alpha]$ is defined to be the discrete measure  on
${\cal HL}_1$\cite{abhay-viqar}.
Thus, a state is normalizable only if it is nonzero on a
countable set of loops.   We may note that this inner product is
introduced only
to define the kinematical quantization.  The physical states, which are
those in the kernal of the constraints, will not be normalizable
with respect to it.

We may now define the operators\cite{gravitons},
\f
\hat{t}^k[\gamma ] \Psi [\vec{\alpha}] \equiv \Psi [\vec{\alpha}
\cup_k \gamma ]
\ff
where
$\vec{\alpha} \cup_1 \gamma \equiv \{ \alpha_1 \cup \gamma , \alpha_2, \alpha_3
\}$ (and
similarly  for $k=2,3$) and
\f
\hat{\tilde{E}}^{ak} (x) \Psi [\vec{\alpha}] \equiv
 \imath \hbar l^2 \Delta^a [x, \alpha_k]  \Psi [\vec{\alpha}]
\ff
The reader may verify that the commutator algebra of these operators
is $i \hbar $ times the corresponding Poisson brackets
and that $\hat{\tilde{E}}^{ai}(x)$ is hermitian and
$\hat{t}^k [\gamma ] $ is unitary
with respect to the inner product (29).  Thus we see that the
reality conditions
are satisfied by this choice of the  inner product.

As in full general relativity\cite{carlolee}, the
diffeomorphism constraint (6) can be
represented by
\f
D(v) \Psi [ \alpha _1, \alpha _2, \alpha _3 ] =  {d\over dt} \Psi
[\phi_t \circ \alpha_1,\phi_t \circ \alpha_2,\phi _t \circ  \alpha _3]
\ff
where $v$ is the vector field which generates the one parameter family of
diffeomorphisms  $\phi_t$.
The diffeomorphism invariant states, which satisfy
\f
D(v) \Psi[\vec{\alpha}] = 0
\ff
are
\f
\Psi[ \alpha_1, \alpha _2, \alpha_3 ] =  \Psi [ L (\alpha_1, \alpha_2,
\alpha_3)]
\ff
where $L(\alpha_1, \alpha_2, \alpha_3) $ are the diffeomorphism
equivalence classes of ${\cal HL}_1^3$ , which we will call
generalized, labled, link classes.

One may also find a large subset of the solution space
of the Hamiltonian constraint (7).  Following methods developed
in\cite{leeted,carlolee}
one easily shows that the solution space to the Hamiltonian
constraint includes the following three types of labled graphs:

1)  All characteristic functions of labled loops with intersections in which
loops of only one type meet.

2)  All superpositions of labled loops of the form
\f
|A  > = A_{i_i ... i_n } \hat{t}^{i_1} [\alpha_1 ] ... \hat{t}^{i_n} [\alpha_n]
|0>
\ff
where $|0>$ is the state that has support (and is equal to unity)
only on trivial
loops and the coefficients $A_{i_i ... i_n }$ are symmetric in the
$i$ and $j$ indices
any time the loops $\alpha_i$ and $\alpha_j$ have a point of
intersection at which
they are non-parallel.

3)  Any intersections in which loops with different labels exchange tangent
vectors is allowed.

It is not presently known if these represent all of the
solutions to the Hamiltonian
constraint, this problem is under investigation.
The
physical state space then consists of functions on $
L_c[ \alpha_1,\alpha_2,\alpha_3 ] $,
which we define to be the
diffeomorphism
equivalence classes of labled knotted graphs consisting of  loops which
satisfy the Hamiltonian constraint at each intersection.
We will call these the
hamiltonian knotted graphs.

As usual, the physical states are not normalizable in the unphysical inner
product (29), and a new inner product needs to be introduced on the
physical state
space.  The most natural candidate for a physical inner product is
\f
<\Psi|\chi>=
\Sigma_{Lc}\bar{\Psi} [L_c] \chi[L_c ]
\ff
where $L_c$ are the Hamiltonian Knotted graphs.   This form
must be considered a conjecture,  which is to be verified by showing that
it satisfies the reality conditions for the physial observables as
described in \cite{poona,gravitons,carlolee,report}. That this may
be the case is suggested from the fact that  it can be constructed
from (29) by a formal
operation in which the amplitude for each loop is normalized by the volume of
its  orbit under diffeomorphisms in the discrete measure $d\mu [ {\alpha}]$.

This completes the Hamiltonian quantization of the system.  The states are
given by the functions on the Hamiltonian knotted graphs, normalizable
under the inner product (36).  As in the full theory, the
key problem remains tthe construction of the  physical observables, and the
verification of the conjecture for the form of the inner product.
In particular,
it is a nontrivial problem to construct the known physical observables
 (23) and (24) on this Hilbert space.  to flesh out the
physical interpretation of the theory.  Finally, we may also note that
the physical meaning of a Hilbert space associated with the Euclidean
quantum theory is obscure.  This problem is certainly connected with the
general problem of time in quantum cosmology\cite{time}.

Finally, it would be interesting to try to extend these methods
to the Minkowskian
case.  The only difference at the Hilbert space level
is that the inner product must now
be chosen to implement the Minkowskian reality conditions.  This is
a non-trivial
problem, as, at least for the nondegenerate sector, the Minkowskian
reality conditions reduce the classical field theory to
a topological field theory with no local degrees of freedom.
To carry out this
program will then be to raise the important issue of the role of
the degenerate
configurations in the quantum theory\footnote{We may note that this
issue arrises
in the $2+1$
case, however there it is resolved by the fact that the
degenerate and nondegenerate
metrics are linked by the gauge transformations
of the theory (See \cite{2+1}).
This is the case for
the $3+1$ theory for some, but not all degenerate metrics.}.

\section{The Euclidean path integral}

We may note that for the Googly theory, as well as for the full theory, the
Hamiltonian and diffeomorphism constraints may be solved using the method of
Capovilla, Dell and Jacobson\cite{CDJ}.
This allows us to construct the measure
for the path integral non-perturbatively, resolving
an issue that has been the
subject of some discussion\cite{measure}.
The path integral may be derived from the Hamiltonian formulation in
the usual way\cite{faddeevslavnov}.  The expression for the
partition function is,
\f
Z=\int [dA_{ai}][d\tilde{E}^{bj}]
e^{-\int d^4 x \tilde{E}^{ai}\dot{A}_{ai}}\prod_x
\delta [C(x)]\delta [C_a (x) ] \delta [ g^i (x) ] \delta (\xi )
\Delta_{F-P}
\ff
where $\xi $ represents seven gauge fixing conditions
per point and $\Delta_{F-P}$ is
the Faddeev-Poppov determinant\cite{faddeevpoppov}.
The solutions to the Hamiltonian
and diffeomorphism constraints are given, just as in the
full theory, by\cite{CDJ},
$A_{ai}$ and a symmetric tracefree matrix field $\phi^i_{\ j}$,
which together determine $\tilde{E}^{ai}$  by
\f
\tilde{E}^{ai} = [\phi^{-1}]^i_{\ j} \tilde{B}^{aj}
\ff
where $\tilde{B}^{aj} = \epsilon^{abc}f_{bc}^j$ and
$[\phi^{-1}]^i_{\ j}= \phi^i_{\ k} \phi^k_{\ j}- 1/2
\delta^i_j (\phi^k_{\ k} )^2 $.
We may note that this solution fails to hold for cases where
$det(\tilde{B}^{ai} ) \neq 0$, but as these are of
measure zero in the original
canonical measure in (40), this should not affect
the path integral.  Using these solutions to
eliminate the Hamiltonian and
diffeomorphism constraints, and introducing $A_0^i$
as a lagrange multiplier to
exponentiate the Gauss's law constraint we find that
\f
Z=\int [dA_{\mu}^i ][d\phi^i_{\ j} ]{\cal J}[A,\phi ]
e^{-\int d^4 x \epsilon^{\mu \nu \alpha \beta }
{}^4f_{\mu \nu i} {}^4f_{\alpha \beta}^j
(\phi^i_{\ k} \phi^k_{\ j}- 1/2 \delta^i_j (\phi^k_{\ k} )^2 )}
 \delta (\xi ) \Delta_{F-P}
\ff
where $A_\mu^i$ is now the spacetime connection and
${\cal J}[A,\phi ]$ is the determinant that comes from
solving the constraints, which is,
\f
{\cal J}[A,\phi ] = \prod_x  det \left [
{\partial \tilde{E}^{ai}(x)  \over \partial \phi^l_{\ m} } \right ] =
\prod_x  det \left [ -( \phi^i_l \phi^m_{\ j}) \tilde{B}^{aj} \right ]
= \prod_x [-(det \tilde{B})^3 (det \phi )^6 ]
\ff
The definition of the path integral is then completed
by specifying a choice of
the seven gauge fixing functions $\xi$, and by introducing
ghosts to exponentiate
the resulting determinants.  The evaluation of this path integral will be
discussed in a future publication, for the present,
we may note the following points.

1)  With a choice of gauge fixing functions which is
linear in $A_a^i$, the path
integral is Gaussian in $A_a^i$, if we use ghosts to exponentiate the
determinants in ${\cal J}[A, \phi ]$ and $\Delta_{F-P}$.
After the Gaussian integral is done one is left
with a path integral over $\phi^i_{\ j}$, and the ghost fields.
As it comes from a theory whose classical solutions are the
self-dual metrics (on which
are propagating the linearized antiself-dual modes),
it seems possible that this expression may
be interesting from a topological point of view,
in view of the role that self-dual
connections have recently played in topology.

2) The path integral (43) is completely diffeomorphism invariant, unlike the
$G \rightarrow 0$ limit of the standard perturbative path
integral.  Thus it seems
a better candidate on which to base a perturbation theory in
$G$ than the standard
perturbative definition, in which the leading order term looses
the gauge symmetry
of the full theory.

3)  The degenerate metrics found in the Ashtekar formulation are
explicitly taken
into account in the formulation of (43), and seem to cause no
particular problems.  This
is interesting on account of the role that the degenerate
metrics play in the
solvable formulation of $2+1$ quantum gravity invented by
Witten\cite{Witten-2+1}.

4)  As the basic variable is a connection, one may attempt to use (43), or its
extension to the full theory, as the starting
point for a Monte-Carlo evaluation of the path integral for quantum
gravity.  The
priniciple technical problem to be faced is that the use of fermionionic
variables seems essential, in order to include correctly the Faddeev-Poppov
determinants that are necessary to correctly handle the
diffeomorphism invariance.

5)  Finally, the action in the exponential (43) is not manifestly bounded.
However, at least in the
asymptotically flat case, in which the Hamiltonian is bouded
from below, the path integral
should be convergent as the measure we have constructed implements
the constraints
that insure positive energy\cite{kristen}.

6)  Essentially the same construction can be used to
construct the path integral
for the full theory.  This, and the issue of the convergence of these
path integrals,  will be discussed elsewhere.

\section{Conclusions}

The googly theory is a reduction of general relativity that
may provide an interesting tool for studying problems in both classical
and quantum gravity.  At the classical level, it is interesting to have a
Hamiltonian system whose solution space is essentially the self-dual
metrics\cite{charles-sd}.  This may play an interesting role in studies of
the geometry and topology of self-dual
spacetimes\cite{selfdual,charles-topology}.
In addition, the speculations that the self-dual
sector may be integrable can now
be pursued within a Hamiltonian framework.

At the quantum level, the googly theory provides a diffeomorphism invariant
quantum field theory that is
non-trivial, but still simpler than quantum gravity. Its quantization
seems to  be interesting in
both the canonical and path integral frameworks.  For example, as more
can be learned about the physical observables in this case it becomes possible
to test the conjecture\cite{book,poona,carlolee}
that the physical inner product is determined by
reality conditions on the physical observables.  Furthermore, if the classical
theory is integrable than  there then ought
to be an infinite dimensional algebra of physical observables, or constants of
the motion.    If they can be constructed, then one may attempt to
define their action in the quantum theory in terms
of the representation defined
here, or more directly via Isham's group quantization program\cite{isham}.

However, what is even more interesting is the possibility
that this theory provides
a starting point for a new definition of perturbative
quantum gravity in which
exact diffeomorphism invariance is respected at every order.
Such a perturbation
theory would be purely quantum mechanical, rather than
semiclassical, in that no
shift of the operators around a purely classical background is
made-an assumption
that is surely suspect at the Planck scale.  Instead, the zero'th
order states and
observables must be exactly diffeomorphism invariant.
This may be expected to have profound consequences for the problem of
the ultra -violet divergences.

\section*{Acknowledgements}

This paper arose out of work in collaboration with Abhay Ashtekar and Carlo
Rovelli, whose influence on the ideas  presented here is
pervasive. I would like to thank them and also Ted Jacobson, Jorge Pullin,
Joseph Samuel and Charles Torre
for discussions about this work.  In addition, I would like to thank the
referee for pointing out an error in an earlier version of this work.
This work was supported by the Natural
Science Foundation under PHY 9012099.


\begin{thebibliography}{99}

\bibitem{abhay}   A. Ashtekar.    {\it New variables for classical and
quantum gravity}.  Physical Review Letters 57, 2244--2247 (1986).
{\it New {H}amiltonian formulation of general relativity},
  Physical Review  D36, 1587--1602 (1987).

\bibitem{book} A. Ashtekar (with invited contributions), {\it New
Perspectives in Canonical Gravity}. Lecture Notes. Bibliopolis, Napoli, Italy,
February 1988.

 \bibitem{poona} A. Ashtekar , {\it Non-perturbative canonical gravity}.
 Lecture notes prepared in collaboration with Ranjeet S. Tate.
(World Scientific
 Books, Singapore,1991).

\bibitem{gravitons}A. Ashtekar, C. Rovelli and L. Smolin
{\it Gravitons and Loops},
Max Planck Institute Munich, Syracuse and Pittsburgh preprint (1991), to
appear in Phys. Rev. D (1991).


\bibitem{roger-googly} R. Penrose, Twistor Newsletter   3 (1976)
12;9 (1979) 9,32

\bibitem{twistor}R. Penrose and W. Rindler, {\it Spinors and spacetime},
Vol. 2, Cambridge University Press, and references contained within.

\bibitem{remarkable}J. Samuel, Syracuse preprint in preparation (1991).

\bibitem{carlolee} C. Rovelli and L. Smolin, {\it Loop representation for
quantum General Relativity}, Nucl. Phys. B133 (1990) 80; Phys. Rev. Lett.
{\bf 61}, 1155 (1988).

\bibitem{gambini} R. Gambini and A. Trias, Phys. Rev. D23 (1981) 553,  Lett.
al Nuovo Cimento 38 (1983) 497, Phys. Rev. Lett. 53 (1984) 2359, Nucl. Phys.
B278 (1986) 436,  D39 (1989) 3127.  R. Gambini, {\it Loop space
representation of quantum general relativity and the group of loops},
preprint University of Montevideo 1990, to appear in Physics Letters
B (1991).  R. Gambini and L. Leal, {\it Loop
space coordinates, linear representations of the diffeomorphism group and
knot invariants} preprint, University of Montevideo, 1991.

\bibitem{action}J. Samuel,   Pramana-J Phys. 28 (1987) L429;
T. Jacobson and L. Smolin, Phys. Lett. B 196 (1987) 39; Class. and Quant.
Grav. 5 (1988) 583.

\bibitem{degenerate}I. M. Bengstrom,
Class. and Quant. Grav. 5 (1988) L139; 7 (1990) 27; preprint 1991;
V. Husain and L. Smolin, Nucl. Phys. B 327 (1989) 205;
L. Smolin, Modern Phys. Lett. A (1989)1091; .

\bibitem{selfdual}A. Ashtekar, T. Jacobson and L. Smolin, Commun.
Math. Phys.

\bibitem{viqar-strong}V. Husain,  Class. and Quant. Grav. 5 (1988) 575.

\bibitem{review} C. Rovelli, {\it The Ashtekar formulation of general
relativity and the loop-representation of quantum gravity.} to appear on
Class. and Quant. Grav. (1991).

\bibitem{maxwell}
A. Ashtekar and C. Rovelli, {\it Quantum Faraday lines; Loop representation
of the Maxwell theory}, Syracuse preprint (1991).

\bibitem{loopym}   C. Rovelli and L. Smolin.   {\it  Loop representation for
lattice gauge theory},   1990 Pittsburgh and Syracuse preprint. B. Bruegmann,
Physical Review D 43 (1991) 566.

\bibitem{2+1} A. Ashtekar, V. Husain, C. Rovelli, J. Samuel and
L. Smolin  {\it
2+1 quantum gravity as a toy model for the 3+1 theory} Class. and Quantum
Grav.  L185-L193 (1989).

\bibitem{leeted}T. Jacobson and L. Smolin, Nucl. Phys. B 299 (1988) 295.

\bibitem{abhay-viqar}D. Rayner, Class. and Quant. Grav. 7 (1990) 651 and
Abhay Ashtekar, personal communication, have suggested the use of this measure.

\bibitem{time} C. Rovelli, {\it Time in quantum gravity: physics beyond the
Schr\"odinger regime}, Universita' di Roma preprint (1989); {\it Time in
quantum gravity: an hypothesis} Physical Review D, January 15th, 1991.



\bibitem{CDJ} R. Capovilla, J. Dell and T. Jacoboson,
Phys. Rev. Lett. 63 (1989)
2325; Class. and Quant. Grav. 8 (1991) 59;
 R. Capovilla, J. Dell, T. Jacoboson and L. Mason, Class. and Quant. Grav.
8 (1991) 41.

\bibitem{measure}V. de Alfaro, S. Fubini and G. Furlan, Nov.
Cim. 57B (1981)227;
K. Fujikawa, Nuclear Physics B 226 (1983) 437; H. Leutwyler,
Phys. Rev. 134 B (1964) 1157;
E. Fradkin and G. Vilkovisky, Phys. Rev. D 8 (1973) 4241.

\bibitem{faddeevslavnov}L.D. Faddeev and Slavnov {\it Gauge
Field Theories: Introduction to the Quantum Theory}
(Benjamin/Cummings,Reading, Mass,1980).

\bibitem{faddeevpoppov}L. D. Faddeev and V.N. Popov,
Phys. Lett. 25 B (1967) 29.

\bibitem{Witten-2+1}E. Witten, Nuclear Physics B 311 (1988) 46.

\bibitem{charles-sd}C. G. Torre, Phys. Lett. B 252 (1990) 242.

\bibitem{charles-topology}C. G. Torre, Phys. Rev. D41 (1990) 3620;
J. Math. Phys.
31 (1990) 2983.

\bibitem{isham}C. J. Isham, in {\it Relativity, Groups and Topology II}, Les
Houches 1983, ed. B. DeWitt and R. Stora (North Holland,Amsterdam,1984);
C. J. Isham and A. C. Kakas, Class. and Quant. Grav. 1 (1984) 621.

\bibitem{kristen}K. Schleich, Phys. Rev. D 36 (1987) 2342; 39 (1989) 2192.

\end{thebibliography}
  \end{document}